\documentclass[pra,aps,twocolumn]{revtex4}
\usepackage{graphicx}
\usepackage{amsmath}

\begin{document}
\title{High-fidelity atomic-state teleportation protocol with
           non-maximally-entangled states}

\author{Grzegorz Chimczak and Ryszard Tana\'s}

\affiliation{Department of Physics, Nonlinear Optics Division, Adam
  Mickiewicz University, 61-614 Pozna\'n, Poland}

\date{\today} \email{chimczak@kielich.amu.edu.pl}

\begin{abstract}
  We propose a protocol of the long-distance atomic
  state teleportation via cavity decay, which allows for high-fidelity
  teleportation even with currently available optical cavities. The
  protocol is based on the  
  scheme proposed by Bose~\emph{et al.} [Phys. Rev. Lett.
  {\textbf{83}}, 5158 (1999)] but with one important modification: it 
  employs non-maximally-entangled states instead of maximally entangled states.
\end{abstract}

\maketitle

\thispagestyle{empty}
\section{Introduction}
Recent years witnessed considerable progress both in theoretical and
experimental quantum information science. The long-range goal in the
field is the realization of quantum networks composed of many nodes and
channels. The present status of the research in the field has been
reviewed in~\cite{kimble08:_quant_inter}. The nodes of the quantum
network require quantum systems that can store quantum information for
sufficiently long time and quantum channels which should allow for fast
transfer of quantum information between the nodes. 
A single atom (or ion) can be considered as a perfect quantum memory ---
qubit can be stored in atomic states even for $10$ s~\cite{langer05_10s}.
Thus, trapped atoms are candidates for being components of quantum
registers or nodes of quantum networks. Fast connections between the
nodes can be realized with photonic qubits which are the best carriers
of quantum information. To transfer quantum information stored in one
node to another node through the photonic channel, it is necessary to
have effective methods for mapping atomic states into field states and
back~\cite{blinov04:_obser_of_entan_between_singl,volz06:_obser_of_entan_of_singl,wilk07:_singl_atom_singl_photon_quant_inter,boozerPRL07_map,choi08:_mappin_photon_entan_into_and}. A number of schemes for
creating entanglement and performing quantum teleportation has been proposed~\cite{cirac97,enk98:_photon_chann_for_quant_commun,cabrillo99,bose,duan_nature,duan:_effic,feng_entanglement,sun04:_atom_photon_entan_gener_and_didtr,chou05:_measur_induc_entan_for_excit,moehring07:_entan_of_singl_atom_quant,yin07:_multiat_and_reson_inter_schem,wu07:_effec_schem_for_gener_clust}.
Next step would be to accomplish the long-distance atomic-state
teleportation mediated by photons, but this task appears to be very challenging
and has not been experimentally achieved yet.

\begin{figure}[htbp]
  \centering
  \includegraphics[width=7cm]{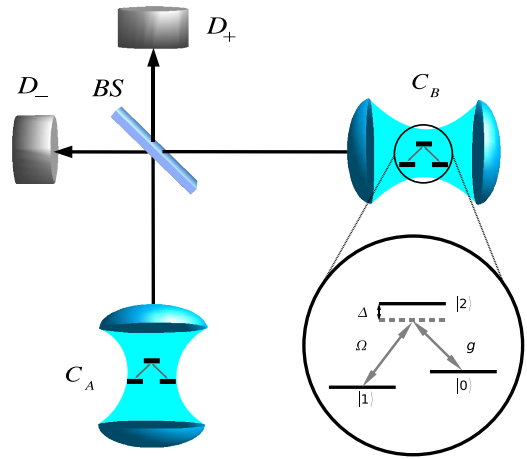}
  \caption{(Color online) The teleportation device and level scheme of the $\Lambda$
  atom interacting with the classical laser field with coupling strength $\Omega$
  and with the quantized cavity mode with the coupling strength $g$.
  Both fields are detuned from the corresponding transition frequencies
  by $\Delta$.}
  \label{fig:device}
\end{figure}

A pretty simple way to complete a long-distance teleportation of
atomic states mediated by photons was proposed by Bose~\emph{et
  al.}~\cite{bose}. Some modifications of this protocol can also be found
in~\cite{browne_entanglement,cho04:_quant_telep_with_atoms_trapp_in_cavit,yu04:_robus_high_fidel_telep_of}. 
The teleportation scheme of Bose~\emph{et~al.}~\cite{bose}  
consist of two atom-cavity systems, a 50:50 beam splitter, and two
detectors as depicted in Fig.~\ref{fig:device}. With this device the
teleportation can be carried out by just performing the joint
detection of both cavities 
fields if, before detection, the sender (Alice) maps the state of
her atom onto the field state of her cavity, and the receiver (Bob) creates the
maximally entangled state of his atom and his cavity field.
Recent progress in technology allows for such state
mapping~\cite{boozerPRL07_map,choi08:_mappin_photon_entan_into_and} 
and performing the joint detection~\cite{legero04}. Creation of the
maximally entangled state of the atom-cavity system also should be
possible with the current technology. However, the Bose~\emph{et~al.}
protocol~\cite{bose} is hardly feasible because the fidelity of state mapping is
drastically reduced by large damping values of the currently available
cavities.

In this paper we propose a modification of the
Bose~\emph{et~al.} scheme~\cite{bose} consisting in 
exploiting, instead of the maximally entangled state, a non-maximally-entangled 
state with the amplitudes chosen
in such a way that the damping factors introduced 
by the state mapping are fully compensated for. With this modification of
the protocol, it should be possible to achieve high teleportation
fidelities even with currently available cavities. The price we have 
to pay for the higher fidelities is a lower probability of success.

\section{Teleportation protocol with non-maximally entangled states}
First, let us present the main idea in a simplified way --- 
comparing it to the standard teleportation protocol~\cite{bennett_tele,nielsen}.
In the standard teleportation protocol Alice has unknown to her
(and to Bob) qubit $|\phi\rangle=\alpha |0\rangle+\beta
|1\rangle$ and one qubit of the Einstein-Podolsky-Rosen (EPR) pair. The second qubit of the
EPR pair is on Bob's site. Suppose, however, that we have the
situation depicted in Fig.~\ref{fig:device}. The state to be teleported
is initially stored in the Alice atom and next is mapped using the laser
to the cavity field qubit, but the mapping is not perfect, and the
initial state is slightly distorted. Let the state of the Alice cavity
field takes the form
\begin{equation}
  \label{eq:1}
  |\phi'\rangle={\cal N}(\alpha|0\rangle+\zeta\beta|1\rangle),
\end{equation}
where ${\cal N}=1/\sqrt{|\alpha|^{2}+|\zeta|^{2}|\beta|^{2}}$ is the
normalization factor and $\zeta$ is a parameter that measures to what
degree the original state has been distorted. If there is no
distortion $\zeta=1$, and the state is just the original state. It is
important that the parameter $\zeta$ does not depend on the original
state (it does not depend on $\alpha$ and $\beta$) but depends solely
on the mapping procedure which is known for both parties of the
protocol. Both parties can agree on the details of the procedure
in advance. Now the question arises: can we use our knowledge of
$\zeta$ to improve the fidelity of the teleported state?   

\begin{figure}[h]
  \centering
  \includegraphics[width=8.6cm]{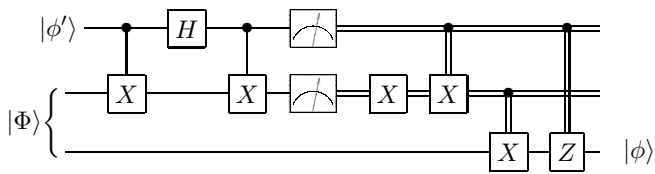}
  \caption{Modified teleportation circuit.}
  \label{fig:teleport}
\end{figure}

The standard teleportation protocol would teleport the distorted
state~[Eq.~\eqref{eq:1}] to Bob. However, if we choose the
non-maximally-entangled state, instead of the maximally entangled state, in the
teleportation protocol, we can correct the imperfections introduced by
the mapping procedure by using a slightly modified
teleportation protocol. The teleportation circuit for this protocol is
illustrated in Fig.~\ref{fig:teleport}. 

The first qubit is the Alice cavity field qubit, the state of which is
initially the state $|\phi'\rangle$ given by equation~Eq.~\eqref{eq:1},
and the state $|\Phi\rangle$ is the non-maximally-entangled state
given by
\begin{equation}
\label{eq:2}
  |\Phi\rangle=a|10\rangle+b|01\rangle.
\end{equation}
The overall initial state is thus
\begin{align}\label{eq:3}
  |\Psi_{0}\rangle&=|\phi'\rangle\otimes|\Phi\rangle=
{\cal N}\left(\alpha|0\rangle+\zeta\beta|1\rangle\right)\otimes\left(a|10\rangle
+b|01\rangle\right)\nonumber\\ 
&={\cal N}\left[\alpha|0\rangle\left(a|10\rangle+b|01\rangle\right)
+\zeta\beta|1\rangle\left(a|10\rangle+b|01\rangle\right)\right].
\end{align}
It is easy to show using the circuit from Fig.~\ref{fig:teleport} that
just before the measurements, the state is given by
\begin{align}\label{eq:4}
  |\Psi_{1}\rangle&=\frac{\cal N}{\sqrt{2}}\left[\vphantom{\frac{\cal N}{\sqrt{2}}}
|00\rangle\left(b\alpha|1\rangle+a\zeta\beta|0\rangle\right)
+|11\rangle\left(b\alpha|1\rangle-a\zeta\beta|0\rangle\right)\right.\nonumber\\
&\left.+|01\rangle\left(a\alpha|0\rangle+b\zeta\beta|1\rangle\right)
+|10\rangle\left(a\alpha|0\rangle-b\zeta\beta|1\rangle\right)
\right].
\end{align}
Now, we see that if we prepare the non-maximally-entangled
state~[Eq.~\eqref{eq:2}] in such a way that $a=\zeta b$ we obtain
\begin{align}
  \label{eq:5}
  |\Psi_{1}\rangle&=\frac{{\cal N}b}{\sqrt{2}}\left[\vphantom{\frac{\cal N}{\sqrt{2}}}
\left[|00\rangle \left(\alpha|1\rangle+\zeta^{2}\beta|0\rangle\right)
+|11\rangle\left(\alpha|1\rangle-\zeta^{2}\beta|0\rangle\right)\right]\right.\nonumber\\
&\left.+\zeta\left[|01\rangle \left(\alpha|0\rangle+\beta|1\rangle\right)
+|10\rangle\left(\alpha|0\rangle-\beta|1\rangle\right)\right]\vphantom{\frac{\cal N}{\sqrt{2}}}
\right].
\end{align}
When Alice performs the measurement on her two qubits, there are two
cases when only one of the detectors registers a 
photon, and the state is projected either to $|01\rangle$ or
$|10\rangle$. Since we assume that the beam splitter is used in the
measuring apparatus, only the two outcomes are considered as successful
because the beam splitter can only distinguish two states from the Bell
basis. The other two outcomes are rejected as unsuccessful. 
Alice next communicate to Bob, using the classical channel, the results
of her measurement (two classical bits), and Bob
applying the postmeasurement operations shown in
Fig.~\ref{fig:teleport}, can recover the original Alice's state
$|\phi\rangle$ with the perfect fidelity. 

Of course, the teleportation scheme depicted in
Fig.~\ref{fig:teleport} works perfectly well as the standard
teleportation protocol when 
the measuring device can distinguish all four Bell states, the
original undistorted state $|\phi\rangle$ is initially on the first
qubit ($\zeta=1$), and the shared entangled state $|\Phi\rangle$ is
the maximally entangled state ($a=b=1/\sqrt{2}$).

\section{Physical model}
In the first stage of teleportation protocol, when Alice has to map
the initial state of her atom $|\phi\rangle=\alpha |0\rangle+\beta
|1\rangle$ onto 
the field state of her cavity and when Bob has to create an entangled state
of his atom and his cavity field, the most important role in the teleportation
protocol play the two atom-cavity systems. Let us first describe them
in more detail. Alice and Bob can change the state of their own
atom-cavity system by switching their lasers on.
When the laser illuminates the atom trapped inside the cavity then the evolution
of the atom-cavity system is governed by the effective non-Hermitian
Hamiltonian ($\hbar=1$ here and in the following),
\begin{eqnarray}
  \label{eq:Hamiltonian0}
  H&=&(\Delta-i\gamma) \sigma_{22} 
  +(\Omega \sigma_{21}+g a \sigma_{20}+ {\rm{H.c.}})
  -i \kappa a^{\dagger} a \, , \nonumber \\
\end{eqnarray}
where $\sigma_{ij}\equiv |i\rangle \langle j|$ denote the atomic flip operators
and $a$ denotes the annihilation operator of the cavity field
mode. One mirror in each cavity is partially transparent to allow for the joint
measurement of the fields leaking out from both cavities. Of course,
the transparency of the mirror leads to a damping of the cavity field
mode. We assume that photons leak out of the cavity at a rate $2\kappa$.
For simplicity, we neglect the spontaneous decay rate of the excited
atomic state $\gamma$. This approximation 
is valid if conditions $\Delta\gg g,\Omega,\gamma$ and $\gamma g^2/\Delta^2$,
$\gamma\Omega^2/\Delta^2\ll\kappa$ are fulfilled~\cite{chimczak02:_effect}.
We can further simplify Hamiltonian~(\ref{eq:Hamiltonian0}) assuming that $\Omega =g$.
Then, after adiabatic elimination of the excited atomic state,
the Hamiltonian takes the form
\begin{eqnarray}
  \label{eq:Hamil1}
  H=-\delta \sigma_{11}-\delta a^{\dagger} a \sigma_{00}
  -(\delta a \sigma_{10} +{\rm{H.c.}})
  -i \kappa a^{\dagger} a \, ,
\end{eqnarray}
where $\delta=g^2/\Delta$. Using Hamiltonian~(\ref{eq:Hamil1})
one can easily get analytical expressions describing evolution of the
initial quantum 
states $|0\rangle_{\textrm{atom}}|0\rangle_{\textrm{mode}}$ 
and $|1\rangle_{\textrm{atom}} |0\rangle_{\textrm{mode}}$. First of
the states experiences no dynamics because there is no operator in
Hamiltonian~(\ref{eq:Hamil1}) which can change  this state. The
evolution of the second state is given by
\begin{eqnarray}
  \label{eq:U}
  e^{-i H t} |10\rangle&=& 
  e^{i \delta t} e^{-\frac{\kappa t}{2}} \big[
  i\,a(t) |01\rangle +b(t) |10\rangle \big] \, , 
\end{eqnarray}
where we abbreviate the atom-cavity state
$|j\rangle_{\textrm{atom}}\otimes|n\rangle_{\textrm{mode}}$ to
$|j n\rangle$ and we use
\begin{eqnarray}
  \label{eq:AiB}
  a(t)&=& 
  \frac{2 \delta}{\Omega_{\kappa}} 
  \sin\Big(\frac{\Omega_{\kappa} t}{2}\Big) \, , \nonumber \\
  b(t)&=&\cos\Big(\frac{\Omega_{\kappa} t}{2}\Big)
  +\frac{\kappa}{\Omega_{\kappa}} 
  \sin\Big(\frac{\Omega_{\kappa} t}{2}\Big) \, , 
\end{eqnarray}
where $\Omega_{\kappa}=\sqrt{4 \delta^2-\kappa^2}$. If the laser is turned
off ($\Omega=0$) then the Hamiltonian takes the form
$H=-\delta a^{\dagger} a \sigma_{00}-i \kappa a^{\dagger} a$
and then the time evolution of the system can be obtained using
the relations
\begin{eqnarray}
  \label{eq:las0}
  e^{-i H t} |10\rangle&=& |10\rangle \, , \nonumber \\
  e^{-i H t} |01\rangle&=& e^{i \delta t} e^{-\kappa t}|01\rangle  \, .
\end{eqnarray}
Equations~(\ref{eq:las0}) are needed to describe evolution of
the device state during the second stage, in which the joint measurement
of both cavities fields is performed. At this stage of the protocol
the most important role play the detectors $D_{+}$
and $D_{-}$ together with the beam splitter $BS$. Registration of the
photon emission by one of the detectors corresponds to the action
of the collapse operator on the joint state of Alice's and Bob's systems.
The collapse operator has the form
\begin{eqnarray}
  \label{eq:C}
  C=\sqrt{\kappa}(a_{A}+i\epsilon a_{B}) \, ,
\end{eqnarray}
where $\epsilon$ is $1$ for photon detection in $D_{+}$ and $-1$ for
photon detection in $D_{-}$.

\section{Teleportation via cavity decay with non-maximally-entangled states}
\begin{figure}[h]
  \centering
  \includegraphics[width=7cm]{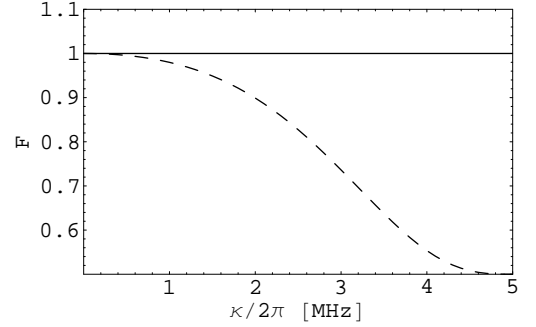}
  \caption{The teleportation fidelity as a function of cavity decay rate $\kappa$
    for the modified protocol (solid line) and the original protocol
    (dashed line) for $(\Delta,g)/(2\pi)=(100,16)$ MHz.}
  \label{fig:fk}
\end{figure}

Now, we can analyze the modified teleportation protocol which makes it
possible to compensate fully for the destructive effect of cavity
decay and, as we believe, it could be realized even with currently available cavities.
The teleportation protocol consists of three
stages, so it is as simple as the original teleportation protocol of
Bose~\emph{et~al.}~\cite{bose}. The three stages are
(A) the preparation stage, (B) the detection stage, and
(C) the recovery stage.
At the beginning of the protocol Alice's atom
is prepared in a state, which is unknown for Alice.
Bob's atom is prepared in the state $|1\rangle_{\textrm{atom}}$.
Initially the field modes of both cavities are empty, so
the states of both atom-cavity systems are given by
\begin{eqnarray}
  \label{eq:p0}
  |\psi\rangle_{A}&=&|\phi\rangle_{\textrm{atom}}\otimes|0\rangle_{\textrm{mode}}\
  =\alpha|00\rangle_{A} +\beta |10\rangle_{A} \, , \\
  \label{eq:p1}
  |\psi\rangle_{B}&=&|10\rangle_{B} \, .
\end{eqnarray}
As we have mentioned above, Alice has to map the state stored in her
atom onto the field state of her cavity in the preparation stage.
She can do it by just turning her laser on for the time
$t_{A}=(2/\Omega_{\kappa})[\pi-\arctan(\Omega_{\kappa}/\kappa)]$~\cite{bose,
chimczak07:_improv_fidel_in_atomic_state}.
After this operation her atom-cavity system is found to be in the state
\begin{eqnarray}
  \label{eq:p3}
  |\widetilde{\psi}\rangle_{A}=\alpha|00\rangle_{A} 
+i e^{i\delta t_{A}} e^{-\kappa t_{A}/2}\beta |01\rangle_{A} \, .
\end{eqnarray}
It is seen that the state mapping is done although
it is imperfect because of the damping factor
$e^{-\kappa t_{A}/2}$. We cannot avoid this damping factor, but
we can show that it is possible to compensate for it. To this aim, in the
modified teleportation protocol, 
Bob creates a non-maximally-entangled state
instead of creating maximally entangled state as it is done in the standard
teleportation protocol. He turns his laser on for time $t_{B}$
changing his system state to 
\begin{eqnarray}
  \label{eq:p4}
  |\widetilde{\psi}\rangle_{B}=e^{-\kappa t_{B}/2} \big[
  i a(t_{B}) |01\rangle_{B} +b(t_{B}) |10\rangle_{B} \big] \, .
\end{eqnarray}
The expression for $t_{B}$ will be given later. Now, we have to
derive the expression for probability that the first stage is successful.
The preparation stage will succeed only under the absence of photon detection
event. Probabilities that no collapse occurs during Alice's and Bob's
operations are given by the squared norms of the state vectors~(\ref{eq:p3})
and~(\ref{eq:p4}), respectively. They are given by
\begin{eqnarray}
  \label{eq:Pa}
  P_{A}&=&|\alpha|^{2} + e^{-\kappa t_{A}} |\beta|^{2} \, , \nonumber \\
  P_{B}&=&e^{-\kappa t_{B}} \big( |a(t_{B})|^2+|b(t_{B})|^2 \big) \, .
\end{eqnarray}
Alice and Bob complete their actions in the same instant of time.
Then they turn the lasers off and the detection stage starts.
Alice during the second stage just waits for a finite time $t_{D}\gg\kappa^{-1}$
registering events of photon detection. This stage and the whole teleportation
protocol is successful when Alice registers one, and only one,
photon. In other cases, when Alice registers no photon or when
she registers two photons, the initial Alice's state is destroyed.
Until the time of photon detection $t_{j}$ the evolution of the state of both
atom-cavity systems is given by~(\ref{eq:las0}), and at time $t_{j}$
both systems states are described by
\begin{eqnarray}
  \label{eq:p7}
  |\widetilde{\psi}(t_{j})\rangle_{A}&=&\frac{1}{\sqrt{P_{A}}}
\big(i e^{i\delta (t_{A}+t_{j})}
e^{-\kappa (t_{A}+2 t_{j})/2}\beta |01\rangle_{A} \nonumber \\
&&+\alpha|00\rangle_{A}\big) \, ,  \\
  |\widetilde{\psi}(t_{j})\rangle_{B}&=&\frac{e^{-\kappa t_{B}/2}}{\sqrt{P_{B}}}
\big( i a(t_{B}) e^{i \delta t_{j}} e^{-\kappa t_{j}} |01\rangle_{B} \nonumber \\
&&+b(t_{B}) |10\rangle_{B} \big) \, .
\end{eqnarray}
The probability of no photon emission before time $t_{j}$ is given
by $P_{D}(t_{j})=P_{A}(t_{j}) P_{B}(t_{j})$, where
\begin{eqnarray}
  \label{eq:p9}
  P_{A}(t_{j})&=&\frac{|\alpha|^2+e^{-\kappa (t_{A}+2 t_{j})} 
  |\beta|^2}{|\alpha|^2+e^{-\kappa t_{A}}|\beta|^2} \, , \nonumber \\
  P_{B}(t_{j})&=&\frac{|a(t_{B})|^2 e^{-\kappa 2 t_{j}}+|b(t_{B})|^2}{|a(t_{B})|^2+|b(t_{B})|^2} \, .
\end{eqnarray}
At time $t_{j}$ one of the detectors registers a photon emission, what corresponds to
the change in the joint state of both atom-cavity systems. After the collapse
the joint state is given by 
$|\widetilde{\phi} (t_{j})\rangle=C |\psi (t_{j})\rangle_{A}\otimes |\psi (t_{j})\rangle_{B}$.
The probability that the collapse occurs in the time interval $t_{j}$
to $t_{j}+dt$ can be calculated from $2\langle \widetilde{\phi}
(t_{j})|\widetilde{\phi} (t_{j})\rangle dt$. 
After the collapse we have to normalize the state
$|\widetilde{\phi} (t_{j})\rangle\to|\phi (t_{j})\rangle$
and then the evolution of the joint state can again be
determined using Eq.~(\ref{eq:las0}), and at the end of
the detection stage, at time $t_{D}$, the state is given by
$|\widetilde{\phi}(t_{D})\rangle=\exp[-i H (t_{D}-t_{j})]|\phi(t_{j})\rangle$.
For $t_{D}\gg\kappa^{-1}$ the normalized joint state can be very well
approximated by
$|\phi(t_{D})\rangle=|00\rangle_{A}\otimes |\psi_{B}(t_{D})\rangle$,
where
\begin{eqnarray}
  \label{eq:f7}
  |\psi_{B}(t_{D})\rangle &=&\frac{e^{i\delta t_{A}} e^{-\kappa t_{A}/2}\beta b(t_{B}) |10\rangle_{B}
+\epsilon \alpha i a(t_{B}) |00\rangle_{B}}
{\sqrt{e^{-\kappa t_{A}}|\beta|^2 |b(t_{B})|^2 + |\alpha|^2 |a(t_{B})|^2}}  \nonumber \\
\end{eqnarray}
From Eq.~(\ref{eq:f7}) it is seen that the unwanted damping factor
$e^{-\kappa t_{A}/2}$ disappears if the condition
\begin{eqnarray}
  \label{eq:f8}
e^{-\kappa t_{A}/2} b(t_{B})&=&a(t_{B})
\end{eqnarray}
is satisfied. We can now give the expression for the time $t_{B}$;
\begin{eqnarray}
  \label{eq:tB}
t_{B}&=&\frac{2}{\Omega_{\kappa}}
\Bigg[\arctan\Bigg(\frac{\Omega_{\kappa} e^{-\kappa t_{A}/2}}{2 \delta -e^{-\kappa t_{A}/2} \kappa}\Bigg)
+n \pi\Bigg]  \, .
\end{eqnarray}
The time given by expression~(\ref{eq:tB}) is the key parameter, which
must be known to Bob to create the non-maximally-entangled state~(\ref{eq:2}).
Time $t_{B}$ is the function of $t_{A}$, so both these times
must be known to Bob. Note, however, that neither time $t_{B}$ 
nor time $t_{A}$ depend on the amplitudes of the teleported state.
If in the preparation stage Bob turns his laser on for time $t_{B}$
then, after the detection stage, the state of his system becomes identical
to the initial unknown Alice's state except for the phase factor
\begin{eqnarray}
  \label{eq:f9}
  |\psi_{B}(t_{D})\rangle &=&\alpha |00\rangle_{B}
  -i\epsilon e^{i\delta t_{A}} \beta |10\rangle_{B} \, .
\end{eqnarray}
Fortunately, the phase factor can be removed using the Zeeman
evolution~\cite{bose}, what Bob performs in the last
stage after receiving classical information about Alice's measurement.
At the end of the whole protocol Bob has the original Alice's
state stored in his atom
$|\phi\rangle=\alpha |0\rangle_{\textrm{atom}}+\beta |1\rangle_{\textrm{atom}}$.
It turns out that the fidelity of this teleportation protocol can be
close to unity even for realistic cavity decay rates. 
Figure~\ref{fig:fk} shows teleportation fidelities of both protocols
(the modified with non-maximally-entangled state and the Bose~\emph{et al.}
protocol) as functions of the cavity decay rate.

\begin{figure}[h]
  \centering
  \includegraphics[width=7cm]{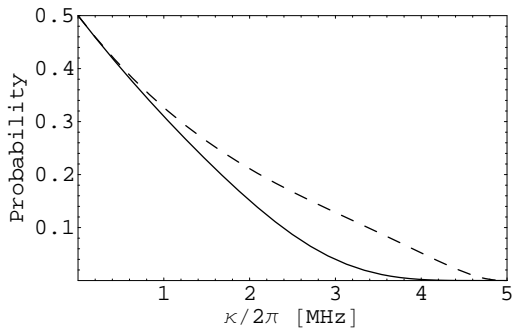}
  \caption{Probability of success versus cavity decay rate $\kappa$
    for the modified protocol (solid line) and the original protocol
    (dashed line) for $(\Delta,g)/(2\pi)=(100,16)$ MHz.}
  \label{fig:4}
\end{figure}

One can see that for real cavity decay rate
$\kappa/2\pi=3.8$~MHz~\cite{boozerPRL07_map} the fidelity of teleported
state is still equal unity while the fidelity of the original protocol
does not exceed the value $2/3$. This result is quite impressive
but one can easily note that the high fidelity is not for free. Since
$P_{A}$ depends on the damping factor, the probability
that the teleportation protocol will be successful is lowered by
the increasing cavity decay rates. Let us now estimate the probability of
success for currently available
cavities. The probability that all stages of the protocol will succeed has
the following form:
\begin{eqnarray}
  \label{eq:Psuk}
P_{\textrm{suc}}&=&P_{A} P_{B}\int_{0}^{t_{D}} P_{A}(t_{j}) P_{B}(t_{j}) 
\langle \widetilde{\phi}(t_{D})|\widetilde{\phi}(t_{D})\rangle \nonumber \\
&&\times 2\langle \widetilde{\phi}(t_{j})|\widetilde{\phi}(t_{j})\rangle dt_{j} \, .
\end{eqnarray}
For $t_{D}\gg\kappa^{-1}$, the probability of success can be very well
approximated by
\begin{eqnarray}
  \label{eq:Psuk8}
P_{\textrm{suc}}&=& e^{-\kappa t_{B}} a(t_{B})^2   \, .
\end{eqnarray}
We can use this simple formula to estimate the value of the success
probability for the experimental parameters of Ref.~\cite{boozerPRL07_map},
i.e., we take $(g,\kappa)/2\pi=(16,3.8)$ MHz. The protocol requires, however,
bigger values of the detuning than that of Ref.~\cite{boozerPRL07_map},
so we take $\Delta/2\pi=100$ MHz. With this set of parameters we get
the probability of success of about $0.005$, and this is the price we
have to pay for getting fidelity close to unity. In Fig.~\ref{fig:4}
we plot the probability of success
as a function of the cavity decay rate $\kappa$ for the modified
protocol and compare it to the corresponding dependence for the
original protocol of Bose~{\em et al.}~\cite{bose} for the parameter
values $(\Delta,g)/(2\pi)=(100,16)$ MHz. It is seen that
the probability of success for the modified protocol goes to zero faster
than that for the original protocol, but it still has considerable
values for realistic decay rates.

Let us now take into account the important imperfection
which is present in all real setups, i.e., finite detection efficiency.
This imperfection is caused by absorption in the mirrors, photon losses
during the propagation between the cavities and the detectors, and
by nonunity detectors efficiency. In Ref.~\cite{mckeever03:_exper}
the overall detection efficiency is only $\eta=0.05$. Therefore,
with such efficiency only a small fraction of all successful runs will
be detected. 
Moreover, the case of two photons emissions will be erroneously
counted as a successful case if only one photon is detected.
Of course, this effect would lead to lowering of the fidelity.
The two-photon case is also very important if detectors cannot
distinguish a single photon from two photons, since both emitted
photons are always collected by the same detector~\cite{legero04},
it is not possible to reject such unsuccessful runs.
If we want to estimate the
real values for the teleportation fidelity and the success probability,
we have to include the efficiency $\eta$ and the two-photon case in
our calculations. 
The probability that two photons will be emitted from the cavities
during the teleportation protocol and only one of them will be
detected in the detection stage is given by
\begin{eqnarray}
  \label{eq:P2em10}
P_{\textrm{2em}}(\eta)&=& |\beta|^2 e^{-\kappa t_{A}} \eta (1-\eta\xi) \, ,
\end{eqnarray}
where $\xi=1$ for photon-number-resolving detectors and
$\xi=1-P_{\textrm{suc}}$ for conventional photon
detectors. This probability depends on the modulus of the amplitude $\beta$
which is in general unknown. Hence, it is necessary to compute
the average probability of two photon emissions taken over all
possible input states. Such an average probability takes the form
\begin{eqnarray}
  \label{eq:P2em11}
\overline{P}_{\textrm{2em}}(\eta)&=&e^{-\kappa t_{A}} \eta (1-\eta\xi)/2  \, .
\end{eqnarray}
The average probability that the measurement will indicate success
is then given by
\begin{eqnarray}
  \label{eq:Peta}
\overline{P}_{\textrm{suc}}(\eta)&=& \eta P_{\textrm{suc}}+\overline{P}_{\textrm{2em}}(\eta) \, .
\end{eqnarray}
In the case of two-photon emissions Bob's atom is in the state $|0\rangle$.
If we cannot reject all runs in which two photons were emitted
then the final state of Bob's atom is a mixture of $|0\rangle$
and $|\phi\rangle$, i.e.,
\begin{eqnarray}
  \label{eq:ro}
\rho&=& \frac{\eta P_{\textrm{suc}} |\phi\rangle\langle\phi| 
+P_{\textrm{2em}}(\eta)|0 \rangle\langle 0|}{\eta P_{\textrm{suc}}+P_{\textrm{2em}}(\eta)} \, .
\end{eqnarray}
We can calculate the average fidelity using the density matrix $\rho$.
The average fidelity of the teleportation protocol is given by
\begin{eqnarray}
  \label{eq:Feta}
\overline{F}(\eta)=1/2+P_{\textrm{suc}}/B-(P_{\textrm{suc}}/B)^2 \ln(1+B/P_{\textrm{suc}}) \, ,
\end{eqnarray}
where $B=\exp(-\kappa t_{A}) (1-\eta\xi)$. It is now evident that there
is one more important feature of the protocol. Expressions~(\ref{eq:Peta})
and~(\ref{eq:Feta}) indicate that for large cavity decay rates,
it is almost irrelevant if the detectors can distinguish a single
photon from two photons. For sufficiently large cavity decay rates
$P_{\textrm{suc}}$ is small, and, therefore,
$\xi$ for conventional detectors is close to unity.
For example, the parameter's regime $(\Delta, \Omega, g, \kappa)/(2\pi)=$
$(100,16,16,3.8)$ MHz leads to $\xi=0.995$.
Hence, possible implementations of the protocol with currently available
cavities do not require detectors with the single-photon resolution.

\section{Numerical results}
\begin{figure}[htbp]
  \centering
  \includegraphics[width=7cm]{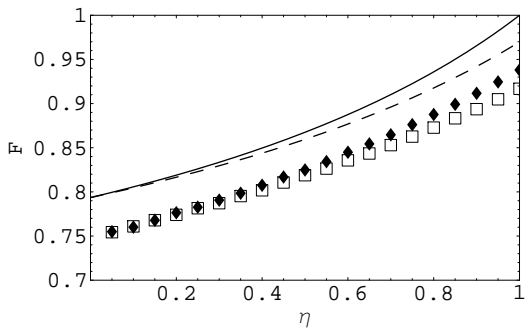}
  \caption{The average fidelity of teleportation given by~Eq.~(\ref{eq:Feta}) for
           photon-number-resolving detectors (solid curve) and for conventional
           single-photon detectors (dashed curve) as a function of the overall detection
           efficiency. The diamonds and open squares refer to numerical results
           for photon-number-resolving and conventional detectors, respectively.
           The numerical results
           include spontaneous emission from excited atomic state and
           dark counts. The parameters are $(\Delta,\Omega,g,\kappa,\gamma)/(2\pi)=(62.5,16,16,4,2.6)$ MHz.}
  \label{fig:Feta}
\end{figure}

\begin{figure}[htbp]
  \centering
  \includegraphics[width=7cm]{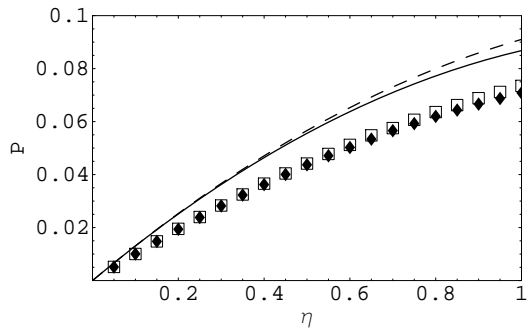}
  \caption{The average probability of success vs the overall detection
           efficiency. For detectors, which are able to distinguish a
           single photon from two photons analytical results given
           by~Eq.~(\ref{eq:Peta}) are presented by the solid curve and
           numerical results are presented by diamonds. The dashed curve
           and open squares correspond to analytical and numerical results,
           respectively, for conventional single-photon detectors. The
           parameters are
           $(\Delta,\Omega,g,\kappa,\gamma)/(2\pi)=(62.5,16,16,4,2.6)$
           MHz.} \
  \label{fig:Peta}
\end{figure}

The analysis of experimental feasibility of this protocol requires
taking into account another imperfection, which is the spontaneous
emission from the atom. We have done it 
using numerical calculations. In the following we present some details
of the calculations.
We have calculated the average fidelity
and the average probability of success using
Hamiltonian~(\ref{eq:Hamiltonian0}) and the quantum trajectories
method~\cite{pleniotrajek,carmichaelksiazka}. Unfortunately,
the evolution of the atom-cavity system is
different than that described by Eq.~(\ref{eq:U}) for
the parameters of Ref.~\cite{boozerPRL07_map}:
the population of the excited state $|2\rangle$ cannot be neglected
and the periodic behavior of the system is lost because of the damping
present in the system. Nevertheless, we can choose parameters
close to that of Ref.~\cite{boozerPRL07_map} for which the average
fidelity of the teleportation protocol is still high.
For the well chosen parameters, times $t_{A}$ and $t_{B}$ should
not be too long as compared to $\kappa^{-1}$ and $\gamma^{-1}$.
If we want to satisfy this condition, we have to set
$\Delta$ to be small enough. Then, however, we get considerable
population of the excited state $|2\rangle$. Fortunately,
this population oscillates, and we can use the fine tuning
technique~\cite{chimczak08:_fine_tunin_of_quant_operat} to minimize
its effect. Applying this technique we have chosen
$(\Delta,\Omega,g,\kappa,\gamma)/(2\pi)=(62.5,16,16,4,2.6)$ MHz.
For these parameters analytical expressions for $t_{A}$ and $t_{B}$
are not precise enough, and therefore we have used numerically optimized
times $t_{A}=0.1058\mu\textrm{s}$ and $t_{B}=0.0131\mu\textrm{s}$
and not too long detection time $t_{D}=4\kappa^{-1}\approx 0.16\mu\textrm{s}$.
The detection time $t_{D}=4\kappa^{-1}$ is long enough to get a quite
high value of the fidelity~\cite{bose}. We do not set longer
times of the detection stage to make the influence of the dark counts
on the protocol negligible. For the dark count rate of
$50~\textrm{s}^{-1}$~\cite{keller04:_contin_gener_of_singl_photon} the mean time between
dark counts ($10$~ms for both detectors) is much larger than the
time window for detection in the protocol $t_{A}+t_{D}\approx
0.3~\mu\textrm{s}$. 
Thus the dark counts can be neglected. Nevertheless, we
have taken them into account in our numerical calculations. Results
obtained from quantum trajectory approach are presented in
Figs.~\ref{fig:Feta} and~\ref{fig:Peta}.

In order to analyze the experimental feasibility of the protocol and
abilities of improving the fidelity, we plot in Fig.~\ref{fig:Feta}
the average fidelity as a function of the overall detection efficiency $\eta$.
As it is evident from the figure, the average fidelity tends to $0.794$
with decreasing $\eta$. So, the average fidelity significantly exceeds
the value $2/3$ even for the real overall detection efficiency $\eta=0.05$.
Of course, the spontaneous decay rate $\gamma/2\pi=2.6$~MHz and dark
counts of $50~\textrm{s}^{-1}$ reduce the average fidelity, but it
is still well above the limit of $2/3$. It is important because the average
fidelity of the teleportation based on classical resources only cannot
exceed this
limit~\cite{riebe04,massar95:_optim_extrac_of_infor_from}. Note that 
the protocol makes it possible to achieve values of the fidelity much
higher than $0.794$. 
In principle, we can obtain the fidelity even close to unity, but it
would require better than currently available 
overall detection efficiencies. The effect of the overall detection
inefficiency on the teleportation protocol is much stronger than that
of other imperfections present in real experimental setups.
Also, the probability of success is lowered by nonideal overall
detection efficiency, as it is evident from Fig.~\ref{fig:Peta}.
The probability of success tends to zero with decreasing $\eta$.
For the currently available efficiency of $0.05$, the success rate has the value
of $0.005$, which means that it takes on average hundreds of runs to get
successful teleportation.
Such small probability of success means that this teleportation protocol
will not have commercial applications for currently available cavities
and detectors. However, this probability is big enough to perform long-distance
teleportation of atomic states and test it. With present day technology
$2000$ trials of protocol that consists state mapping stage and detection
stage last $360$ ms~\cite{boozerPRL07_map} only. Therefore all data required
can be collected in a reasonable time.

From Figs.~\ref{fig:Feta} and~\ref{fig:Peta} it is also seen that
the expensive photon-number-resolving detectors are not necessary for
the parameters used in our computations. For $\eta$ close to unity, there is only
a small difference between the fidelity obtained with the assumption
that the detectors have the ability to distinguish a single photon
from two photons and the fidelity obtained with assumption that
the detectors have not such ability. For the real overall detection efficiency
$\eta=0.05$ the difference is indistinguishable.

\section{Experimental feasibility of the protocol}
Finally, we shortly discuss the realizability of our teleportation protocol.
As mentioned above, almost all parameters used in our computations are feasible
with current technology. The only parameter the value of which may
be demanding for present technology is the detuning. In our numerical
calculations we have chosen $\Delta/(2\pi)=62.5$~MHz, which is the
value six times greater than that of Ref.~\cite{boozerPRL07_map}. 
Moreover, so far we have assumed that the laser pulses have rectangular
shapes. This assumption
makes it possible to examine the proposed teleportation protocol
analytically and numerically. However, the shortest
rising time of such pulse has duration $100$~ns~\cite{boozerPRL07_map,
mckeever04:_single_photon}. Therefore, real pulses that are approximately
rectangular cannot be shorter than $1~\mu\textrm{s}$. The pulses
duration times used in our numerical calculations are much shorter:
$t_{A}=0.1058\mu\textrm{s}$ 
and $t_{B}=0.0131\mu\textrm{s}$. So, it is rather unrealistic to implement 
experimentally the protocol in its present form. Nevertheless,
the protocol can be easily adapted for using other shapes of laser pulses.
All what is actually needed to complete this teleportation protocol is
the ability to perform the state mapping
\begin{eqnarray}
  \label{eq:map}
\alpha|00\rangle+\beta|10\rangle&\rightarrow&
\alpha|00\rangle+e^{-\kappa t/2}\beta|01\rangle\, ,
\end{eqnarray}
and the ability to generate the non-maximally-entangled state
\begin{eqnarray}
  \label{eq:spl}
|10\rangle&\rightarrow& a(t) |01\rangle+b(t)|10\rangle\, ,
\end{eqnarray}
with small $|a(t)|^2$.
First of these operations have already been demonstrated
experimentally~\cite{boozerPRL07_map}. The second operation
can be achieved with short Gaussian pulses.

\section{Conclusions}
In conclusion, we have presented a modified protocol
that, in principle, should allow for atomic-state teleportation via cavity decay
using currently available optical cavities. We have shown that
the destructive influence of large cavity decay on the fidelity of
teleported state can be minimized by using in the teleportation
protocol the non-maximally-entangled state instead of the maximally
entangled state. This happens despite the fact that both of them
separately lead to lowering of the
teleportation fidelity~\cite{ozdemir07_teleport}. 
Advantage of using non-maximally-entangled states
has been indicated also for other quantum information
protocols~\cite{modlawska08:_nonmax_entan_states_can_be}.

We have also shown that there are two other distinguishing features of
the protocol presented here which make it easier to implement experimentally.
First is the possibility of using conventional single-photon
detectors instead of the photon-number-resolving
detectors. Second is the average fidelity exceeding
the limit $2/3$ even for very small values of the overall
detection efficiency. However, the high fidelities of teleported
states for real cavities can be achieved with the protocol at the
expense of accepting low success rates.

\section*{ACKNOWLEDGMENT}
This work was supported by the Polish research network LFPPI.


\end{document}